\documentclass[letterpaper, 10 pt, conference]{ieeeconf}  

\IEEEoverridecommandlockouts                              
\usepackage{graphics} 
\usepackage{epsfig} 
\usepackage{times} 
\usepackage{amsmath} 
\usepackage{amssymb}  

\usepackage{amsthm}

\usepackage[compress]{cite}
\let\labelindent\relax
\usepackage{enumitem}
\usepackage{mathtools}
\usepackage[dvipsnames]{xcolor}
\usepackage{mathrsfs}
\usepackage{placeins}
\usepackage{url}

\usepackage{graphicx}
\usepackage{subcaption}
\usepackage{fancyhdr}
\fancypagestyle{firstpage}{\fancyhf{}\fancyhead[C]{\raisebox{22pt}[0pt][0pt]{\parbox{\textwidth}{\centering\scriptsize \copyright~2026 IEEE. Personal use of this material is permitted. Permission from IEEE must be obtained for all other uses,\\in any current or future media, including reprinting/republishing this material for advertising or promotional purposes,\\creating new collective works, for resale or redistribution to servers or lists, or reuse of any copyrighted component of this work in other works.}}}}
\newcommand{\R}{\mathbb{R}}
\newcommand{\N}{\mathbb{N}}
\newcommand{\Rpos}{\mathbb{R}_{\geq 0}}
\newcommand{\Rspos}{\mathbb{R}_{> 0}}
\newcommand{\CalT}{\mathcal{T}}

\newcommand{\tderiv}[2]{\frac{d #1}{d #2}}
\newcommand{\pderiv}[2]{\frac{\partial #1}{\partial #2}}
\newcommand{\norm}[1]{\left\|#1\right\|}

\newcommand{\abs}[1]{\left|#1\right|}

\newcommand{\lbo}[2]{\mathcal{L} \left(#1 ; #2\right)}

\newcommand{\diff}[1]{\mathscr{D}(#1)}
\newcommand{\bddiff}[3]{\mathscr{D}^{#2}_{[#1]}(#3)}

\newcommand{\bdiff}[2]{\mathscr{D}_{[\mathcal{T} \times#1]}(#2)}

\newcommand{\kbdiff}[2]{\mathscr{D}^k_{[\N \times#1]}(#2)}

\newcommand{\nom}[1]{\bar{#1}}
\newcommand{\nEnv}{\nom{\Omega}}
\newcommand{\nPhi}{\nom{\varphi}}

\newtheorem{theorem}{Theorem}
\newtheorem{lemma}{Lemma}

\newtheorem{remark}{Remark}
\newtheorem{corollary}{Corollary}
\newtheorem{definition}{Definition}

\title{\LARGE \bf
Time-Varying Coverage via Diffeomorphic Coverage Control
}

\author{Joshua Keene $\qquad$ Chris Manzie $\qquad$ Airlie Chapman $\qquad$ Peter M. Dower
\thanks{All authors are with the University of Melbourne. This research is partially supported by the Australian Government and the University of Melbourne Ingenium Program. (Emails: jckeene@student.unimelb.edu.au, \{manziec, airlie.chapman, pdower\}@unimelb.edu.au.)}
}%

\begin{document}

\maketitle
\thispagestyle{firstpage}
\pagestyle{empty}

\begin{abstract}
\looseness-1
This paper addresses a structured class of time-varying coverage control in which region–density evolution is parameterised by a time-varying diffeomorphism acting on a static region–density pair. 
We refer to this setting as diffeomorphic coverage control. Such trajectories arise naturally in mass-conserving transport flows and in density-driven formation control.
We propose decentralised diffeomorphic coverage controllers for both continuous- and discrete-time dynamics that achieve time-varying coverage by solving a related time-invariant coverage problem in a nominal domain. 
The resulting controllers are asymptotically stable with respect to a variant of a popular coverage metric under corresponding time-varying regions and densities. 
Further analysis shows that diffeomorphic coverage corresponds to a time-varying non-uniform coverage problem over a non-Euclidean metric. 
Simulations demonstrate effectiveness of the proposed control strategy under affine diffeomorphisms in a formation control task, achieving performance comparable to a popular centralised coverage control law. 
Coverage performance is also demonstrated under non-affine diffeomorphisms.

\end{abstract}
\section{Introduction}
Voronoi coverage control algorithms have gained significant attention due to their suitability for spatial coordination tasks in multi-agent systems, such as sensor and surveillance networks \cite{cortes2004coverage, Kennedy_Thesis}.
These methods, first demonstrated in \cite{cortes2004coverage} via a continuous-time extension of Lloyd’s algorithm \cite{lloyd_least_1982}, disperse agents over a region to minimise a spatial objective function.
This dispersion can be shaped using a density function which typically encodes relative environmental importance \cite{cortes2004coverage, pimenta_simultaneous_2010}.
The controller in \cite{cortes2004coverage} is decentralised, stable, and robust to agent addition and removal, making it well suited to real-world deployments.

Time-varying coverage control extends the standard problem setting to time-dependent regions and density functions, enabling applications where desired agent dispersion may be time-dependent, e.g. sensing time-varying event distributions \cite{Kennedy_Thesis, cortes_coverage_2002}, collaborative target tracking \cite{pimenta_simultaneous_2010}, or user-defined formation control and human-swarm interaction \cite{lee_multirobot_2015, Xu_nonconvex}. 

To date, time-varying coverage control has proved challenging. Distributed  control laws (in the Voronoi sense \cite{cortes_spatially-distributed_2005}) in \cite{cortes_coverage_2002, pimenta_simultaneous_2010, keene_2024, Zamani_GMM} modify gradient-descent dynamics of \cite{cortes2004coverage} with additional feed forward terms to compensate for region and density dynamics. 
However, closed-loop trajectory existence under these control laws can be difficult to verify \cite{pimenta_simultaneous_2010} and stability certificates generally require restrictive conditions such as non-trivial closed-loop invariance \cite{cortes_coverage_2002} or asymptotically stationary regions and densities \cite{keene_2024}.

Centralised time-varying coverage controllers offer stronger stability guarantees. 
These control laws achieve exponential convergence to local partition centroids under time-varying densities \cite{lee_multirobot_2015, diaz-mercado_humanswarm_2017} and regions \cite{Xu_nonconvex}, and are later generalised to a family of quasi-Newton flow controllers that rely on a state-dependent inverse Hessian of the coverage objective \cite{cortes2004coverage, keene_acc}. 
However, scalability is limited by the need for a centralised, state-dependent matrix inversion, which complicates trajectory existence and introduces numerical difficulties in parts of the state space \cite{keene_acc}.
Designing a scalable, decentralised time-varying coverage controller with stability guarantees under general region and density dynamics therefore remains an open problem.

\looseness=-1
This paper advances this goal by developing a framework for time-varying coverage in which region and density evolution are parameterised by a sufficiently smooth time-varying diffeomorphism applied to a static region–density pair, a setting we term \textit{diffeomorphic} coverage control. 
An example of such a region-density trajectory is given in \cite{teruel_distributed_2019} where a uniformly-weighted region evolves via an affine transformation in a discrete-time, time-varying coverage task. 
We generalise this setting to simultaneous region and density evolution under general time-varying diffeomorphisms. 
This generalisation captures a larger class of region-density trajectories, most notably those of mass-conserving transport flows \cite{Moser, BB00}, supporting broad applicability.

Examples of diffeomorphic coverage control already appear in the literature. 
In \cite{Leonard_cartogram}, a cartogram, a distortion-minimising diffeomorphism, is used to transform a non-uniform coverage problem into a near-Euclidean one.
The Euclidean coverage solution is then mapped back into the original domain to achieve non-uniform coverage \cite{Leonard_cartogram}. 
The approach in \cite{Xu_nonconvex} achieves non-convex coverage over star-shaped regions by similar means through constructing a diffeomorphism that renders the transformed domain convex.

This work proposes a diffeomorphic coverage solution in both the continuous- and discrete-time settings.
The proposed controllers are decentralised and demonstrate asymptotic stability certificates by leveraging stability guarantees from existing static coverage laws \cite{cortes2004coverage, lloyd_least_1982}. 
Furthermore, we establish a connection between time-varying diffeomorphic coverage and non-uniform coverage over time-varying metrics, an extension to the setting in \cite{Leonard_cartogram}. 
As an example, the continuous-time controller is applied to a density-driven formation control task where formations are specified by time-varying region-density trajectory rather than by explicit graphs in a setting we term \textit{amorphous formation control}.

\looseness=-1
In terms of organisation, Section \ref{sec:Time_Varying_Coverage} reviews the time-varying coverage problem, followed by the development of diffeomorphic coverage control in Section \ref{sec:Diffeomorphic_Coverage_Control}. Section \ref{sec:induced_metric_structure} establishes equivalence between diffeomorphic coverage and coverage over a non-uniform metric. Examples demonstrating the efficacy of the approach are provided in Section \ref{sec:Examples} with concluding remarks in Section \ref{sec:Conclusion}.

\looseness=-1
Throughout, $\mathbb{N}, \mathbb{R}, \Rpos, \Rspos$ denote the sets of natural, real, non-negative, and strictly positive real numbers, respectively, while $\mathbb{R}^b, b \in \mathbb{N},$ denotes Euclidean space. For $B \subset \R^b$, let $\mathcal{K}^B$ denote the set of all convex subsets of $B$. For $A \subset \R^a$ let $\mathscr{AC}(A;B), \mathscr{C}^1(A;B), \lbo{\R^a}{B}$ denote the sets of absolutely continuous maps, continuously-differentiable maps, and bounded linear operators from $\R^a$ to $B$, respectively. Further denote the set of continuously differentiable maps with bounded derivatives as $ \mathscr{C}_b^1(A;B) \subset  \mathscr{C}^1(A;B)$. For a Fr\'echet differentiable map $f:A \to B$ the Fr\'echet derivative at $x_0 \in A$ is denoted $D_xf(x_0)=\pderiv{f}{x}(x_0) \in \lbo{\R^a}{B}$ and the corresponding gradient as $\nabla_x f(x_0) \in \lbo{B}{\R^a}$ such that, for all $h \in \R^a$, $D_xf(x_0) [h] = \pderiv{f}{x}(x_0) h = \nabla_x f(x_0)^\top h$. Operator notation $\partial_t \doteq \pderiv{}{t}$ distinguishes partial time ($t$) derivatives. 

\section{The Time-Varying Coverage Problem}
\label{sec:Time_Varying_Coverage}
\looseness-1
Time-varying coverage control concerns the dispersion of $n\in\N$ agents over a coverage region and density function that evolve over time $t\in\CalT\doteq[0,\infty)$. Let a convex, bounded polytope $\Xi\subset\R^2$ denote the environment. Agents disperse over a convex region $\Omega_t\subseteq\Xi$, varying in time via $\Omega_{(\cdot)}:\CalT\to\mathcal{K}^\Xi$ with $\Omega_t$ Lipschitz in $t$ with respect to the Hausdorff metric, i.e. the velocity of the boundary is finite. The density is described by a bounded map $\phi_t:\Xi\to\Rspos$ that is continuously differentiable in $t$. Typically $\phi_t$ represents evolving areas of interest, or may prescribe user-defined agent formations in $\Xi$, see \cite{diaz-mercado_humanswarm_2017}. The region-density pair trajectory $t \mapsto (\Omega_t,\phi_t)$ fully specifies the time-varying coverage task.

\looseness-1
Denote the collection of agent indices as $\mathcal{I} \doteq \{1,2,...,n\}$. The collective agent state is denoted $P=(p^1, p^2, ... , p^n) \in \mathscr{P} \doteq (\R^2)^{n}$, where $p^i$ is the position of the \textit{i\textsuperscript{th}} agent. Collective agent trajectories through time are denoted by the map $t \mapsto \mathcal{P}_t:\mathcal{T} \to  \mathscr{P}$ such that $\mathcal{P}_{t} \in \mathscr{P}$ is the collective agent state at time $t \in \mathcal{T}$. By an abuse of notation, $p^i_t \in \R^2$ may be used to denote the position of the \textit{i\textsuperscript{th}} agent at time $t$.

\looseness=-1
A popular coverage quality metric for a time-invariant region-density pair $(\Omega, \phi)$, subject to a distance function $f:\R^2 \times \R^2 \to \Rpos$, is the coverage energy $W[\Omega, \phi, f] : \mathscr{P} \to \Rspos$, given by
\begin{equation} 
    \label{eq:W(P)}
   W[\Omega, \phi, f](P) \doteq  \frac{1}{2} \int_{\Omega} \min_{i \in \mathcal{I}} f(p^i, q)  \ \phi(q)\ dq.
\end{equation} 

Let $\mathcal{S} \doteq \{P\in \mathscr{P} \ | \ \exists (i,j) \in \mathcal{I}^2, \ i \neq j,  \textrm{s.t.} \ p^i = p^j\}$, i.e. $\mathcal{S}$ is the set of collective agent positions with collocated agents. For non-collocated agents a more convenient form of $W[\Omega, \phi, f]$ is the locational cost \cite{cortes2004coverage} $\mathcal{H}[\Omega, \phi, f]: \mathscr{P} \backslash \mathcal{S} \to \Rspos$, defined as
\begin{equation} 
    \label{eq:calH(P)}
   \mathcal{H}[\Omega, \phi, f](P) \doteq \frac{1}{2}\sum_{i \in \mathcal{I}}  \int_{V^i[\Omega, f](P)} f(p^i, q) \ \phi(q)\ dq,
\end{equation} 
where $V[\Omega, f](P) \doteq \{V^1[\Omega, f](P), ... , V^n[\Omega, f](P)\}$ is the region-bounded $f$-metric Voronoi partition \cite{du_centroidal_1999} generated by agent positions $P$, defined element-wise by

\noindent
\begin{align*}
	V^i[\Omega, f](P) & \doteq \{ q \in \Omega \ | \ f(p^i,q) \leq f(p^j,q)\  \forall j \in \mathcal{I}\}.
\end{align*}
The default distance function is taken as the squared Euclidean metric, $f_e(a,b) \doteq \|a-b\|_2^2$, i.e. $f$ omission implies the use of \(f_e\) such that $ \mathcal{H}[\Omega, \phi] \doteq \mathcal{H}[\Omega, \phi, f_e]$. The Voronoi partition induces a network topology on agents. The collection of Voronoi-neighbours of agent $i$ is denoted by the set of agents sharing a partition boundary, i.e. \mbox{$\mathcal{N}^i[\Omega](P) \doteq \{j \in \mathcal{I}\backslash \{i\} \ | \ V^i[\Omega](P) \cap V^j[\Omega](P) \neq \emptyset \}$.} 

The locational cost in \eqref{eq:calH(P)} is naturally extended to the time-varying region-density pair $(\Omega_t, \phi_t)$  \cite{Xu_nonconvex, keene_2024} 
\begin{align}
    \label{eq:H_t(P)}
    H_t(P) \doteq \mathcal{H}[\Omega_t, \phi_t](P).
\end{align}

Formally, time-varying coverage control aims to design an agent trajectory $t \mapsto \mathcal{P}_t^* \in \mathscr{AC}(\CalT; \mathscr{P} \backslash \mathcal{S})$ such that \eqref{eq:H_t(P)} is minimised pointwise in time, i.e. for all $t \in \CalT$,
\begin{align*}
    H_t(\mathcal{P}_t^*) = \min_{P \in \mathscr{P}\backslash \mathcal{S}} H_t(P) = \min_{P \in \mathscr{P}\backslash \mathcal{S}} \mathcal{H}[\Omega_t, \phi_t](P).
\end{align*}

Stationary points of \eqref{eq:calH(P)} are identified using a first order optimality condition. Remarkably, application of the Leibniz integral rule \cite{du_centroidal_1999, Kennedy_Thesis} to \eqref{eq:H_t(P)} renders gradients which are \textit{decentralised in the sense of Voronoi}, given by  \cite{cortes2004coverage}
\begin{align*}
    \nabla_{p^i} H_t(P) = m^i[\Omega_t, \phi_t](P) \left( p^i - c^i[\Omega_t, \phi_t](P)\right),
\end{align*}
where $m^i, c^i$ correspond to the mass and centroid of the \textit{i\textsuperscript{th}} agent's Voronoi partition defined as, 
\begin{gather}
    \label{eq:centroid}
    \begin{aligned}
         m^i[\Omega_t, \phi_t](P) & \doteq \int_{V^i[\Omega_t](P)} \phi_t(q) \ dq ,\\
         c^i[\Omega_t, \phi_t](P) & \doteq \frac{1}{m^i[\Omega_t, \phi_t](P)} \int_{V^i[\Omega_t](P)} q \ \phi_t(q) \ dq.
    \end{aligned}
\end{gather}
At $t\in\CalT$ a stationary point of \eqref{eq:H_t(P)} satisfies $p^i = c^i[\Omega_t,\phi_t](P)$ for all $i\in\mathcal I$. The set of all agent positions $P$ satisfying this condition for the region–density pair $(\Omega,\phi)$ is denoted $\mathcal{C}[\Omega,\phi] \subset \mathscr{P}$. For $P \in \mathcal{C}[\Omega,\phi]$ the partition $V[\Omega](P)$ is known as a \textit{Centroidal Voronoi Tessellation} (CVT).

Many coverage controllers minimise $H_t$ by driving agents toward and tracking a CVT trajectory, enforcing the first-order optimality condition $\norm{\nabla_P H_t(P)} = 0$ \cite{cortes2004coverage, keene_2024, lee_multirobot_2015, keene_acc}. In the time-varying setting, decentralised gradient-descent methods \cite{keene_2024, cortes_coverage_2002} require restrictive assumptions and generally lack stability guarantees to time-varying CVTs. Conversely, Newton-flow controllers \cite{lee_multirobot_2015, keene_acc} provide stronger stability certificates but require centralised computation. Moreover, non-convexity of $H_t$ may manifest in CVTs undergoing saddle-node bifurcations along trajectories \cite{Kennedy_Thesis}, thereby degrading the numerical conditioning of corresponding Newton-flow methods. Hence, it is desirable to design a decentralised control law that does not suffer from such numerical degradation when tracking CVT trajectories that may undergo saddle-node bifurcations.

\section{Diffeomorphic Coverage Control}
\label{sec:Diffeomorphic_Coverage_Control}
\subsection{Diffeomorphisms}
\looseness=-1
This work considers a structured class of time-varying coverage control where $t \mapsto (\Omega_t,\phi_t)$ is parameterised by a sufficiently regular time-varying spatial diffeomorphism. A map $\Phi:\R^d \! \to \R^d$ is called a diffeomorphism if it is bijective, i.e. $\Phi^{-1}$ exists, and $\Phi,\Phi^{-1} \in \mathscr{C}^1(\R^d; \R^d)$. The set of all such diffeomorphisms is denoted $\diff{\R^d}$. Bijectivity of $\Phi \in \diff{\R^d}$ implies that, for all $x \in \R^d$,
\begin{align}
    \label{eq:diff_id1}
    D_x\Phi^{-1}(x)  & = [D_{x} \Phi(\Phi^{-1}(x))]^{-1}.
\end{align}
\looseness=-1
By an abuse of notation, the action of $\Phi$ on a subset $\nom{\mathcal{A}} \subset \R^d$ is denoted $\Phi(\nom{\mathcal{A}} ) \doteq \{\Phi(\nom{x}) \ | \ \nom{x} \in \nom{\mathcal{A}}  \} \subset \R^d$. Likewise for $m \in \N$, the $\Phi$-image of $\nom{Q} \doteq [\nom{q}^1, .. , \nom{q}^m]\in (\R^d)^m$ is defined element-wise as $\Phi(\nom{Q}) \doteq [\Phi(\nom{q}^1), .. , \Phi(\nom{q}^m)] \subset (\R^d)^m$. 

The set of diffeomorphisms with deformation bounded by the constant $b < \infty$ over $\nom{\mathcal{A}} \subset \R^d$ is denoted $\bddiff{\nom{\mathcal{A}} }{b}{\R^d} \subset \diff{\R^d}$ such that $\Phi \in \bddiff{\nom{\mathcal{A}} }{b}{\R^d}$ if 
\begin{align*}
    \sup_{\nom{x} \in \nom{\mathcal{A}} } \left\{\norm{D_x \Phi(\nom{x})} + \norm{D_x\Phi^{-1}(\Phi(\nom{x}))}\right\} \leq b. 
\end{align*}
Moreover, $\Phi \in \bddiff{\nom{\mathcal{A}} }{b}{\R^d}$ implies the following.
\begin{lemma}[Diffeomorphism Induced Metric Equivalence]
    \label{cor:Metric_Equiv}
    \looseness=-1
    For $\Phi \in \bddiff{\nom{\mathcal{A}} }{b}{\R^d}$, there exists $d[\Phi]: \R^d \times \R^d \to \Rpos$ such that for all $\nom{x}, \nom{y}\in \nom{\mathcal{A}} $, $\norm{\Phi(\nom{x})-\Phi(\nom{y})}_2 = d[\Phi](\nom{x},\nom{y})$ where
    \begin{equation}
        \label{eq:induced_metric}
        \begin{aligned}
        d[\Phi](x,y) & \doteq \norm{g[\Phi](x,y) (x-y)}_2, \\
         g[\Phi](x,y) & \doteq \left[\int_{0}^1 D_x\Phi(y + \tau(x-y)) d\tau \right].
        \end{aligned}
    \end{equation}
    Moreover, the Euclidean and $d[\Phi]$ metrics are equivalent such that for all $\nom{x},\nom{y} \in \nom{\mathcal{A}} $,
    \begin{align*}
        b^{-1} \norm{\nom{x}-\nom{y}}_2 \leq d[\Phi](\nom{x},\nom{y}) \leq b \norm{\nom{x}-\nom{y}}_2.
    \end{align*}
\end{lemma}
\begin{proof}
    The definition of $d[\Phi]$ follows by the mean value theorem \cite{baggett_functional} noting $D_x\Phi(x)$ exists for all $x \in \nom{\mathcal{A}} $. Metric equivalence follows from \eqref{eq:diff_id1} and the definition of $\bddiff{\nom{\mathcal{A}} }{b}{\R^d}$.
\end{proof}
More specifically, we consider the set of sufficiently regular \textit{time-dependent} diffeomorphisms, as defined below.
\begin{definition}[Regular time-dependent diffeomorphisms]
\label{def:regular_diffeomorphism}
Let $\bdiff{\nom{\mathcal{A}} }{\R^d}$ be the collection of maps $\Psi$ such that: (i) $\Psi \in \mathscr{C}^1_b(\mathcal{T}\times\nom{\mathcal{A}} \ ;\mathbb{R}^d)$; (ii) there exists a finite deformation bound $C<\infty$ such that $\Psi_t \doteq \Psi(t, \cdot) \in \bddiff{\nom{\mathcal{A}} }{C}{\R^d}$ for all $t \in \mathcal{T}$.
\end{definition}

Bijectivity of $\Psi \in \bdiff{\nom{\mathcal{A}}}{\R^d}$ implies that, for all $t \in \CalT, \nom{x} \in \R^d$,
\begin{align}
    \label{eq:diff_id2}
    \partial_t \Psi^{-1}_t[\Psi_t(\nom{x})]  &= -D_x \Psi^{-1}_t[\Psi_t(\nom{x})] \partial_t \Psi_t(\nom{x}).
\end{align}

\subsection{Diffeomorphic Region-Density Trajectories}
We restrict attention to time-varying coverage control over region-density trajectories $t \mapsto (\Omega_t, \phi_t)$ that are \textit{diffeomorphic}, i.e., generated by the action of $\Psi \in \bdiff{\nEnv}{\R^2}$ on a fixed region-density pair $(\nEnv, \nPhi)$, defined as follows.

\begin{definition}[Diffeomorphic Region-Density Trajectory] \label{def:diffeomorphic_region-density}
A region-density trajectory $t \mapsto (\Omega_t, \phi_t)$ is said to be \textit{diffeomorphic} if there exists:
\begin{enumerate}[label=(\roman*)]
    \item a bounded and convex polytope $\nEnv \subset \R^2$;
    \item a strictly positive and finite map $\nPhi:\nEnv \to \Rspos$;
    \item a time-dependent diffeomorphism $\Psi \in \bdiff{\nEnv}{\R^2}$,
\end{enumerate} 
such that $(\Omega_t, \phi_t) = \Gamma[\Psi_t](\nEnv, \nPhi)$ for all $t \in \CalT$ where,
\begin{align}
    \label{eq:diffeo_region_density}
    \Gamma[\Phi](\nEnv, \nPhi) \doteq (\Phi(\nEnv), \Phi_\#\nPhi),
\end{align}
and $\Phi_\#\nPhi$ denotes the mass-conserving pushforward of $\nPhi$ over $\Phi$ given by
\begin{align}
    \label{eq:pushforward_density}
    \Phi_\#\nPhi(q) \doteq \nPhi(\Phi^{-1}(q)) \abs{\text{\emph{det}}(D_{x} \Phi^{-1}(q))}.
\end{align}
\end{definition}

The restriction $\Psi \in \bdiff{\nEnv}{\R^2}$ ensures regularity of $(\Omega_t, \phi_t)$ that is typical in time-varying coverage control. Deformation bounds on $\Psi$ ensure that $\phi_t$ is strictly positive and finite and that $\Omega_t$ is bounded for all $t \in \mathcal{T}$. Likewise, derivative bounds on $\Psi$ ensure that $\Omega_t$ and $\phi_t$ both evolve continuously with bounded derivatives.
\begin{remark}
    \looseness-1
    The structure imposed in \eqref{eq:diffeo_region_density} captures many useful classes of region-density trajectories. One natural example is coverage-enabled formation control, see Section \ref{sec:AmorphousFormationControl}. More generally, any density trajectory $t \mapsto \phi_t$ evolving under a mass-conserving transport process admits the representation \eqref{eq:diffeo_region_density} where $\Psi_t$ is the corresponding \textit{Lagrangian} flow map \cite{BB00}. Such densities are common in control and optimisation and are central to optimal transport theory. However, reconstructing $\Psi$ from a given $\phi$ can be challenging \cite{BB00, Moser}. This work considers the setting where $\Psi$ is known explicitly. \hfill{$\square$}
\end{remark}

\subsection{Stability under Regular Time-Varying Diffeomorphisms}
\label{sec:General_Stab_section}
In this work we use dynamics that solve a problem in a time-invariant ''nominal'' domain, $\nom{\mathcal{A}} \subset \R^d$ (the \textit{nominal system}), to address a related problem in a ''physical'' domain evolving via $\Psi \in \bdiff{\nom{\mathcal{A}}}{\R^d}$, $\mathcal{A}_t \doteq \Psi_t(\nom{\mathcal{A}})$ (the \textit{physical system}). The bar notation $\nom{(\cdot)}$ is used to denote quantities associated with the nominal system. Necessary nominal system regularity is codified by the following definition.
\begin{definition}[Regular Nominal System] \label{def:NominalSystem}
\looseness=-1 The map $\nom{U}:\R^d \to \R^d$ is said to generate a regular nominal system over $\nom{\mathcal{A}} \subset \R^d$ if the following holds:
    \begin{enumerate}[label=(\roman*)]
        \item $\nom{U}$ is uniformly bounded over $\nom{\mathcal{A}}$; 
        \item for all $\nom{x}_0 \in \nom{\mathcal{A}}$ there exists a unique $t \mapsto \nom{\xi}_t[\nom{x}_0]$ such that, for all $t \in \CalT$,
        \begin{align}
            \label{eq:nominal_sys}
            \textstyle\tderiv{}{t}\nom{\xi}_t[\nom{x}_0] = \nom{U}(\nom{\xi}_t[\nom{x}_0]), \quad \nom{\xi}_0[\nom{x}_0] = \nom{x}_0;
        \end{align}
        \item $t \mapsto \nom{\xi}_t[\nom{x}_0]$ is invariant to $\nom{\mathcal{A}}$ for all $\nom{x}_0 \in \nom{\mathcal{A}}$.
    \end{enumerate}
\end{definition}
We seek dynamics that generate behaviour in the physical system that is qualitatively equivalent to that of the nominal system. The existence of such dynamics, which generate an \textit{equivalent physical system}, follows by the following lemma.
\begin{lemma}[Equivalent Physical System]
    \label{lem:continuous_equiv_systems}
    \looseness=-1\
    Let $\nom{U}$ generate a regular nominal system over $\nom{\mathcal{A}} \subset \R^d$ per Definition \ref{def:NominalSystem}. Given $\Psi \in \bdiff{\nom{\mathcal{A}} }{\R^d}$, $x_0 \in \Psi_0(\nom{\mathcal{A}})$, a solution to the initial value problem
    \begin{equation}
        \label{eq:induced_dynamics}
        \begin{aligned}
            \textstyle\tderiv{}{t}\xi_t[x_0]  &= U_t[\Psi, \nom{U}](\xi_t[x_0]), \qquad \xi_0[x_0] = x_0,\\
            U_t[\Psi, \nom{U}](x) &\doteq \partial_t \Psi_t\big(\Psi_t^{-1}(x)\big) \\
            &\qquad +  D_x \Psi_t\big(\Psi_t^{-1}(x)\big) \, \nom{U}\big(\Psi_t^{-1}(x)\big),
    \end{aligned}
    \end{equation}
    
    exists and is unique. Moreover, $\xi_t[x_0] \in \Psi_t(\nom{\mathcal{A}})$ and 
    \begin{align}
        \label{eq:equivalence}
        \xi_t[x_0] = \Psi_t \left( \nom{\xi}_t[\Psi^{-1}_0(x_0)]\right),
    \end{align}
    for all $t \geq 0$ where $\nom{\xi}_{(\cdot)}$ is a trajectory of the nominal system.
\end{lemma}
\begin{proof}
    See Appendix \ref{append:cont_proofs}.
\end{proof}
Moreover, equivalent physical system trajectories satisfying \eqref{eq:induced_dynamics} inherit asymptotic stability of nominal system trajectories per the following theorem.
\begin{theorem}[Physical System Asymptotic Stability]
    \label{theorem:continuous_time_stability}
    \looseness=-1
    Let $\nom{U}$ generate a regular nominal system over $\nom{\mathcal{A}} \subset \R^d$ per Definition \ref{def:NominalSystem}. 
    For $\Psi \in \bdiff{\nom{\mathcal{A}} }{\R^d}, \ x_0 \in \Psi_0(\nom{\mathcal{A}})$, let $t \mapsto \xi_t[x_0]$ denote the solution to \eqref{eq:induced_dynamics}.
    Suppose there exists a continuously differentiable function $\nom{V}: \R^d\to \R$ such that $\nom{V}(\nom{x}) \geq 0$ and $\nabla_{x}\nom{V}(\nom{x})^\top \nom{U}(\nom{x}) \leq 0$ holds for all $\nom{x} \in \nom{\mathcal{A}} $. 
    Then $t \mapsto \xi_t[x_0]$ is asymptotically stable and $\lim_{t\to\infty}\textrm{dist}(\xi_t[x_0], \Psi_t(\eta[\nom{U}])) =0$ where $\eta[\nom{U}] \subseteq M[\nom{V}, \nom{U}]$ is largest $\nom{U}$-invariant subset of $M[\nom{V}, \nom{U}]$ defined as
    \begin{align*}
        M[\nom{V}, \nom{U}] \doteq \{\nom{x} \in \nom{\mathcal{A}} \ | \  \nabla_{x}\nom{V}(\nom{x})^\top \nom{U}(\nom{x}) = 0\}.
    \end{align*}
\end{theorem}
\begin{proof}
    See Appendix \ref{append:cont_proofs}.
\end{proof}

\subsection{Diffeomorphic Coverage Control}
\begin{figure}
    \centering
    \includegraphics[width=0.95\linewidth]{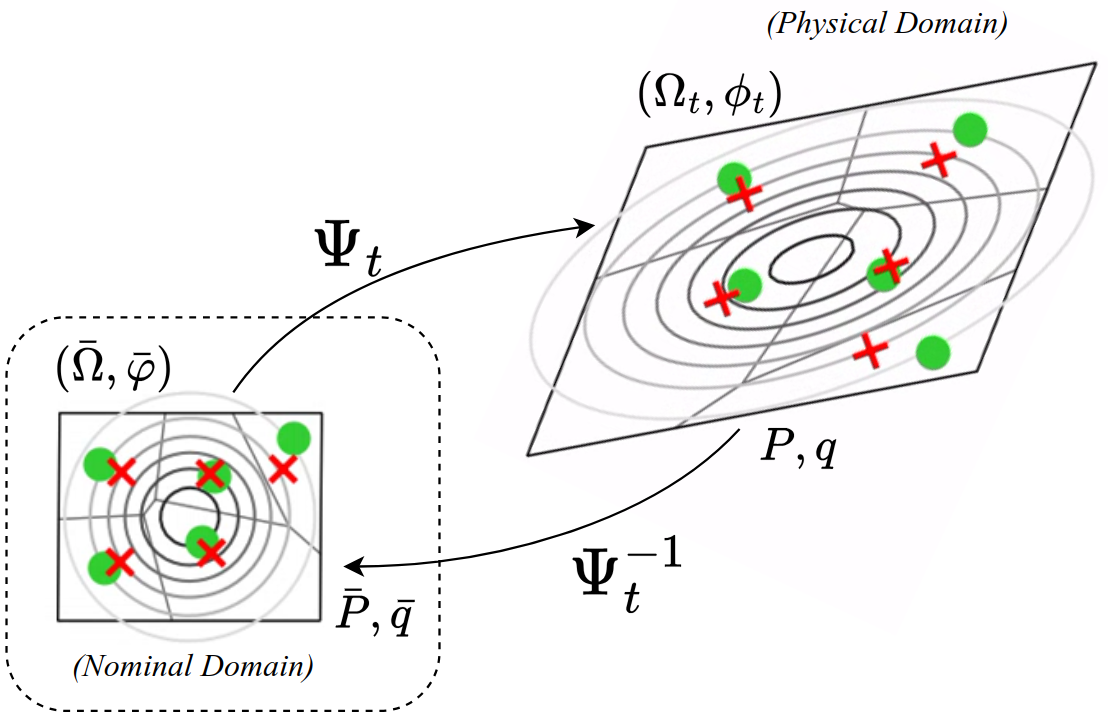}
    \caption{Diffeomorphic coverage control with \textit{nominal} and \textit{physical} domains. Circles, crosses, contours, and lines denote agent and centroid positions, density function level sets, and partition boundaries respectively.}
    \label{fig:diffeomorphic_coverage}
    \vspace{-5mm}
\end{figure}

When $t \mapsto(\Omega_t, \phi_t)$ is diffeomorphic with respect to $(\nEnv,\nPhi)$ and $\Psi \in \bdiff{\nEnv}{\R^2}$, the time-varying coverage problem over $\Omega_t$ in the physical domain is related to a static coverage problem over $\nEnv$ in a nominal domain by the transformation $\Psi$, as illustrated in Figure \ref{fig:diffeomorphic_coverage}. 
We refer to this general setting as \textit{diffeomorphic coverage control}. Diffeomorphic coverage is then a specialisation of the general case in Section \ref{sec:General_Stab_section} and admits a simplified control architecture that directly leverages an existing static coverage control law and inherits corresponding stability certificates. 

\looseness=-1
For the time-invariant region-density pair $(\nEnv, \nPhi)$ and $k > 0$, an asymptotically stable coverage control law in the nominal domain is $\nom{U}[\nEnv, \nPhi]: (\nEnv)^n \backslash \mathcal{S} \to (\R^2)^n$ given by \cite{cortes2004coverage}
\begin{equation}
    \label{eq:cortes}
    \begin{aligned}
    \nom{U}[\nEnv, \nPhi](\nom{P}) & \doteq \left[\nom{u}^i[\nEnv, \nPhi](\nom{P})\right]_{i \in \mathcal{I}}, \\
    \nom{u}^i[\nEnv, \nPhi](\nom{P}) & \doteq -k(p^i - c^i[\nEnv, \nPhi](\nom{P})),
    \end{aligned}
\end{equation}
where $c^i[\nEnv, \nPhi]$ as per \eqref{eq:centroid}. The dynamics in \eqref{eq:cortes} are distributed \textit{in the sense of Voronoi} and local dynamics $u^i[\nEnv, \nPhi]$ only depend on local Voronoi-neighbours $\mathcal{N}^i[\nEnv](\nom{P})$ \cite{cortes_spatially-distributed_2005}. 

A \textit{diffeomorphic} coverage control law over $t \mapsto (\Omega_t, \phi_t)$ in the physical domain then follows from the specialisation of \eqref{eq:induced_dynamics} per the nominal coverage dynamics in \eqref{eq:cortes}, yielding
\begin{equation}
    \label{eq:diffeomorphic_controller}
    \begin{aligned}
    U_t[\Psi, \nom{U}[\nEnv, \nPhi]](P) & = \left[u_t^i[\Psi, \nom{u}^i[\nEnv, \nPhi]](\Psi^{-1}_t(P))\right]_{i \in \mathcal{I}}, \\
    u_t^i[\Psi, \nom{u}](\nom{P}) & \doteq \partial_t \Psi_t(\nom{p}^i) + D_x \Psi_t(\nom{p}^i) \ \nom{u}(\nom{P}),
    \end{aligned}
\end{equation}
where $\nom{u}^i$ as per \eqref{eq:cortes} and agent-separability of \eqref{eq:diffeomorphic_controller} follows from \eqref{eq:cortes}. Corresponding asymptotic stability certificates are established by the following corollary of Theorem \ref{theorem:continuous_time_stability}.
\begin{corollary}[Diffeomorphic Coverage Control]
    \label{prop:diffeo_coverage_control}
    \looseness=-1
    Suppose $t \mapsto (\Omega_{t}, \phi_{t})$ is diffeomorphic with respect to $(\nEnv, \nPhi)$ and $\Psi \in \bdiff{\nEnv}{\R^2}$ as per Definition \ref{def:diffeomorphic_region-density}.
    Then, for any $P_0 \in (\Psi_0(\nEnv))^n \backslash \mathcal{S}$, the closed-loop trajectory generated by $U_t$ in \eqref{eq:diffeomorphic_controller} satisfying
    \begin{align*}
        \textstyle \tderiv{}{t}{\mathcal{P}}_t[P_0] = U_t[\Psi, \nom{U}[\nEnv, \nPhi]](\mathcal{P}_t[P_0]), \qquad \mathcal{P}_0[P_0] = P_0,
    \end{align*}
    exists and is unique. Moreover, $\mathcal{P}_t[P_0] \in \Omega_t$ for all $t \in \CalT$, and $t \mapsto \mathcal{P}_t[P_0]$ is asymptotically stable and approaches $\Psi_t ({\mathcal{C}[\nEnv, \nPhi]})$ as $t \to \infty$. In addition, if $\mathcal{C}[\nEnv, \nPhi]$ is finite, then there exists $C^* \in \mathcal{C}[\nEnv, \nPhi]$ such that 
    \begin{align*}
        \lim_{t \to \infty} \norm{\mathcal{P}_t[P_0] -\Psi_t(C^*)} = 0.
    \end{align*}
\end{corollary}
\begin{proof}
    \looseness=-1
    The dynamics in \eqref{eq:cortes} generate a regular nominal system in $\nEnv^n \backslash \mathcal{S}$ per Definition \ref{def:NominalSystem} \cite{cortes_spatially-distributed_2005}. Existence, uniqueness, and invariance of $t \mapsto \mathcal{P}_t[P_0]$ to $\Omega_t$ follows by Lemma \ref{lem:continuous_equiv_systems}. The proof then follows by Theorem \ref{theorem:continuous_time_stability} using the nominal-domain locational cost, $\mathcal{H}[\nEnv, \nPhi]$ in \eqref{eq:calH(P)}, as a candidate function. By \cite{cortes2004coverage}, $\mathcal{H}[\nEnv, \nPhi]$, is decrescent under dynamics $\nom{U}[\nEnv, \nPhi]$ in \eqref{eq:cortes} with the largest invariant subset being the set of CVTs $\mathcal{C}[\nEnv, \nPhi]$. Therefore, $\mathcal{P}_t[P_0]$ approaches $\Psi_t ({\mathcal{C}[\nEnv, \nPhi]})$ asymptotically. The final assertion follows from \cite[Corollary 1.2]{cortes2004coverage}.
\end{proof}
\looseness=-1
Decentralised Voronoi partition construction via half-plane intersection \cite{cortes2004coverage} remains valid in the physical domain due to the induced metric equivalence in Lemma~\ref{cor:Metric_Equiv}, together with the finite deformation bound $C$  (Definition \ref{def:regular_diffeomorphism}). This guarantees correspondence between sensing radii in the nominal and physical domains. Agent positions measured in the physical domain can therefore be used to directly construct the euclidean Voronoi partitions in the nominal domain. However, larger deformation bounds may necessitate larger sensing radii in the physical domain as per Lemma~\ref{cor:Metric_Equiv}.
\begin{remark}
    \looseness-1
    The diffeomorphic coverage control law \eqref{eq:diffeomorphic_controller} is \emph{decentralised} in the physical domain, making it suitable for large-scale systems, and generates closed-loop trajectories that asymptotically converge to trajectories of nominal-domain CVTs evolving by $\Psi_t$.  Moreover, the diffeomorphic coverage setting and control law in \eqref{eq:diffeomorphic_controller} naturally extend to problems outside of time-varying coverage under appropriate diffeomorphism construction, e.g. non-uniform and non-convex coverage problems~\cite{Leonard_cartogram, xu_formation}.  Note, implementing \eqref{eq:diffeomorphic_controller} requires $\Psi_t$, together with $\partial_t \Psi_t$ and $D_x \Psi_t$, to be known a priori. This is consistent with other time-varying coverage controllers that assume a priori knowledge of $\phi_t, \dot{\phi}_t, \Omega_t$ for all $t \in \CalT$ \cite{lee_multirobot_2015, keene_acc} and estimating these inputs from sampled information remains an open research direction.\hfill{$\square$}
\end{remark}
An analogous controller for discrete-time coverage is deferred to Appendix \ref{sec:append_diffeom_lloyd} for brevity. The proposed \textit{diffeomorphic} Lloyd descent is a time-varying generalisation to the well-known Lloyd Algorithm \cite{lloyd_least_1982} and may prove useful in accommodating more complex dynamics via hierarchal control strategies, or in applications beyond typical coverage such as dynamic mesh generation \cite{du_centroidal_1999}. Stability results follow by analogous arguments used in this section.  

\section{Induced Metrics in Diffeomorphic Coverage}
\label{sec:induced_metric_structure}
Diffeomorphic coverage admits a natural interpretation as coverage over a related curved space. The nominal-domain locational cost $\mathcal{H}[\nEnv, \nPhi]$ underpins stability analysis of the proposed controllers. Motivated by this role, we define the \textit{diffeomorphic locational cost} 
$\nom{H}_t : \mathscr{P} \backslash \mathcal{S} \to \mathbb{R}_{>0}$ 
as the physical-domain representation of $\mathcal{H}[\nEnv, \nPhi]$, i.e.
\begin{equation}
    \label{eq:diffeo_loc_cost}
    \begin{aligned}
        \nom{H}_t(P) & \doteq \mathcal{H}[\nEnv, \nPhi] \left(\Psi_t^{-1}(P)\right).
    \end{aligned}
\end{equation}
$\nom{H}_t$ is distinct from $H_t$ in \eqref{eq:H_t(P)}, with time-dependence in \eqref{eq:diffeo_loc_cost} injected by $t \mapsto\Psi_t^{-1}$ in contrast to injection by $t \mapsto (\Omega_t, \phi_t)$ in \eqref{eq:H_t(P)}. Notably, the following lemma shows that \eqref{eq:diffeo_loc_cost} can be re-cast as a time-varying locational cost in the physical-domain with a $\Psi$-induced distance function.

\begin{lemma}
    \label{prop:metric_structure}
     Let $t \mapsto (\Omega_{t}, \phi_{t})$ be diffeomorphic with respect to $(\nEnv, \nPhi)$ and $\Psi \in \bdiff{\nEnv}{\R^2}$ as per Definition \ref{def:diffeomorphic_region-density}. Then, with  $f[\Phi] \doteq [d[\Phi]]^2$ per \eqref{eq:induced_metric},
     \begin{align*}
         \nom{H}_t(P) = \mathcal{H}[\Omega_t, \phi_t, f[\Psi^{-1}_t]](P).
     \end{align*}
\end{lemma}
\begin{proof}
    First note that by \eqref{eq:pushforward_density}, 
    \begin{align*}
        \phi_t(q) \doteq \nPhi(\Psi^{-1}_t(q))  \abs{\textrm{det}(D_x\Psi^{-1}_t(q))}.
    \end{align*}
    Define $\tilde{V}^i_t(P) \doteq \Psi_t(V^i[\nEnv](\Psi_t^{-1}(P)))$. Applying the change of variables $\nom{q} = \Psi^{-1}_t(q)$ in \eqref{eq:diffeo_loc_cost}, noting the role of the Jacobian and the definition of $d[\Psi_t^{-1}]$ in \eqref{eq:induced_metric}, gives
    \begin{align*}
        \nom{H}_t(P) & = \sum_{i \in \mathcal{I}} \int_{V^i[\nEnv](\Psi_t^{-1}(P))} \norm{\Psi_t^{-1}(p^i)-\nom{q}}^2 \nPhi(\nom{q}) \ d\nom{q} \\
        & = \sum_{i \in \mathcal{I}} \int_{\tilde{V}^i_t(P)} \left[d[\Psi_t^{-1}](p^i,q)\right]^2 \phi_t(q) \ dq.
    \end{align*}
    Moreover, $\tilde{V}^i_t(P) = V^i[\Omega_t, f[\Psi^{-1}_t]](P)$ by similar arguments, yielding the assertion.
\end{proof}

\begin{remark}
    \looseness=-1
    While the present work focuses on time-varying coverage control, Lemma \ref{prop:metric_structure} illustrates that the proposed problem setting is a time-varying generalisation of the non-uniform coverage setting in \cite{Leonard_cartogram}. This generalisation is reflected in the additional time-dependent feed forward term in \eqref{eq:induced_dynamics} when compared to corresponding dynamics in \cite{Leonard_cartogram}. This suggests that diffeomorphic coverage control may also have utility under \textit{time-varying} non-uniform metrics in a manner similar to \cite{Leonard_cartogram}. \hfill{$\square$}
\end{remark}

\section{Examples}
\label{sec:Examples}
\subsection{Amorphous Formation Control}
\label{sec:AmorphousFormationControl}
\looseness-1
A motivating application of diffeomorphic coverage is \textit{amorphous formation control} where an \textit{amorphous} formation is defined locally by a region–density pair $(\nEnv, \nPhi)$ with formation motion then encoded via a time-varying diffeomorphism. Unlike many formation control algorithms, amorphous formation control does not impose a rigid graph structure, motivating our use of the term ``amorphous''. 

\looseness=-1
Amorphous formation control via diffeomorphic coverage offers several advantages over graph-based methods. First, formation definitions via region-density pairs are inherently scalable, independent of agent quantity. Agent–node assignment is also unnecessary, handled implicitly by the coverage law without secondary auctioning or allocation. Furthermore, the agent-agnostic architecture is robust to changing network compositions under agent addition or removal.

In this example we consider amorphous formation control when the local formation, defined via $(\nEnv, \nPhi)$ satisfying \emph{(i), (ii)} in Definition \ref{def:diffeomorphic_region-density}, undergoes translation, $d_{(\cdot)} \in \mathscr{C}^1_b(\mathcal{T};\R^2)$, rotation, $\theta_{(\cdot)} \in \mathscr{C}^1_b(\mathcal{T};\R)$, and uniform scaling $\alpha_{(\cdot)} \in \mathscr{C}^1_b(\mathcal{T};[\underline{\alpha}, \overline{\alpha}])$, $0<\underline{\alpha}\leq\overline{\alpha}<\infty$. For convenience, denote $z_{(\cdot)} \doteq [d_{(\cdot)}^\top, \theta_{(\cdot)}, \alpha_{(\cdot)}]^\top$. The corresponding diffeomorphism of interest is then the similar transform $\breve{\Psi} = \breve{\Psi}[z_{(\cdot)}] \in \bdiff{\nEnv}{\R^2}$ defined as, 
\begin{align*}
    \breve{\Psi}[z] (q)& \doteq \alpha \ R(\theta) \  q + d,  \quad 
    R(\theta)  \doteq 
    \begin{bmatrix} 
    \cos \theta & - \sin\theta \\
    \sin \theta  & \cos \theta
    \end{bmatrix},
\end{align*}
\looseness=-1
for all $q \in \R^2$. Functionally, $d_{(\cdot)}, \theta_{(\cdot)}, \alpha_{(\cdot)}$ define formation position, orientation, and scale with respect to the external environment respectively and offer simplified and separable trajectory definition. Moreover, for all $p \in \nEnv, t \in \CalT$ and $\dot{z}_t = \tderiv{}{t}z_t$, derivatives of $\breve{\Psi}[z_t]$ are given by
\begin{align*}
    \partial_t \breve{\Psi}[z_t](\nom{p})=v[z_t, \dot{z}_t](\nom{p}), \qquad D_x \breve{\Psi}[z_t](\nom{p}) = \alpha_t R(\theta_t), \\
     v[z, \dot{z}](\nom{p})  \doteq \dot{d} + R(\theta) \left(\dot{\alpha} \ \mathbb{I}_{2\times 2} + \alpha \ \dot{\theta} \begin{bmatrix}
        0 & -1 \\
        1 & 0
    \end{bmatrix} \right)\nom{p}.
\end{align*}
Agent dynamics follow from the diffeomorphic coverage control law in \eqref{eq:diffeomorphic_controller} specialised to $\Psi \doteq \breve{\Psi}[z_{(\cdot)}]$, i.e.
\begin{align*}
    \dot{\mathcal{P}}_t \! = \! \!\left[u_t^i[\breve{\Psi}[z_{(\cdot)}], \nom{u}^i[\nEnv, \nPhi]](\breve{\Psi}^{-1}[z_{t}](\mathcal{P}_t))\right]_{i \in \mathcal{I}}. 
\end{align*}
Moreover, the structure of $\breve{\Psi}$ simplifies agent dynamics $u_t^i$,
\begin{align*}
    u_t^i[\breve{\Psi}[z_{(\cdot)}], \nom{u}^i[\nEnv, \nPhi]](\nom{P}) & = u^i[z_t, \dot{z}_t](\nom{P}),
\end{align*}
where, for $\nom{u}^i[\nEnv, \nPhi]$ as per \eqref{eq:cortes},
\begin{align}
    \label{eq:formation_controller}
    u^i[z, \dot{z}](\nom{P}) & \doteq  v[z, \dot{z}](\nom{p}^i) + \alpha \ R(\theta) \  \nom{u}^i[\nEnv, \nPhi](\nom{P}).
\end{align}

\begin{remark}
    If $\alpha_{(\cdot)} = 1$, $\breve{\Psi}[z]$ is a rigid-body transformation and $u^i[z, \dot{z}]$ may be interpreted as a ``leader-follower'' coverage controller. Here, follower agents cover $(\nEnv, \nPhi)$ along trajectories defined by a leader's position and heading $d_t, \theta_t$ while explicitly compensating for dynamics induced by the leader's velocity $\dot{d}_t$ and yaw-rate $\dot{\theta}_t$. \hfill{$\square$}
\end{remark}

The performance of the diffeomorphic coverage control law $u^i[z, \dot{z}]$ is demonstrated by simulation. A static region-density pair $(\nEnv, \nPhi)$ is defined as
\begin{align*}
    \nEnv \doteq [-8, 8]^2, \quad
    \nPhi(q) \doteq \mathcal{N}\left(q \ ;\begin{bmatrix} 0 \\ 0
    \end{bmatrix}, \begin{bmatrix} 15 & 11 \\ 11 & 15
    \end{bmatrix}\right),
\end{align*}
where $\mathcal{N}(\cdot; \mu, \Sigma)$ denotes a normalised multi-variate Gaussian probability density function with mean $\mu \in \R^2$ and covariance $\Sigma \in \R^{2 \times 2}$. 
The simulation occurs over $t \in [0, 15]$ and consists of seven agents aiming to cover the amorphous formation defined in the physical domain by the region-density pair $(\Omega_t, \phi_t) \doteq \Gamma[\breve{\Psi}[z_t]](\nEnv, \nPhi)$, with corresponding formation trajectory $z_t \doteq [d_t^\top, \theta_t, \alpha_t]^\top$ given by
\begin{align*}
    d_t \doteq [5t, 0]^\top, \quad \theta_t \doteq \frac{-\pi}{45} t^2,  \quad 
    \alpha_t \doteq 1 + 0.6 \sin \left(\frac{8 \pi}{225}t^2 \right).
\end{align*}
$(\Omega_t, \phi_t)$ features significant translation, rotation, and scaling that increase in speed over the duration of the simulation. 

\looseness=-1
The coverage performance of $u^i[z,\dot{z}]$ in \eqref{eq:formation_controller} is compared with three existing controllers from the literature. $U_C \doteq \nom{U}[\Omega_t,\phi_t]$ \eqref{eq:cortes} denotes the coverage law of \cite{cortes2004coverage} directly applied to the physical domain, $[U_t]_K$ is the distributed gradient-descent controller from \cite{keene_2024}, and $[U_t]_L$ the centralised controller of \cite{lee_multirobot_2015}. Simulations use Euler integration with a $0.005\text{s}$ time step and identical initial conditions. Feedback gains of $k=12$ are used for $u^i[z,\dot{z}]$ and $[U_t]_L$, while $k=18$ for $U_C$ and $[U_t]_K$ is required to ensure agents remain within $\Omega_t$ over the simulation.
\begin{figure}
    \centering
    \includegraphics[width=1\linewidth]{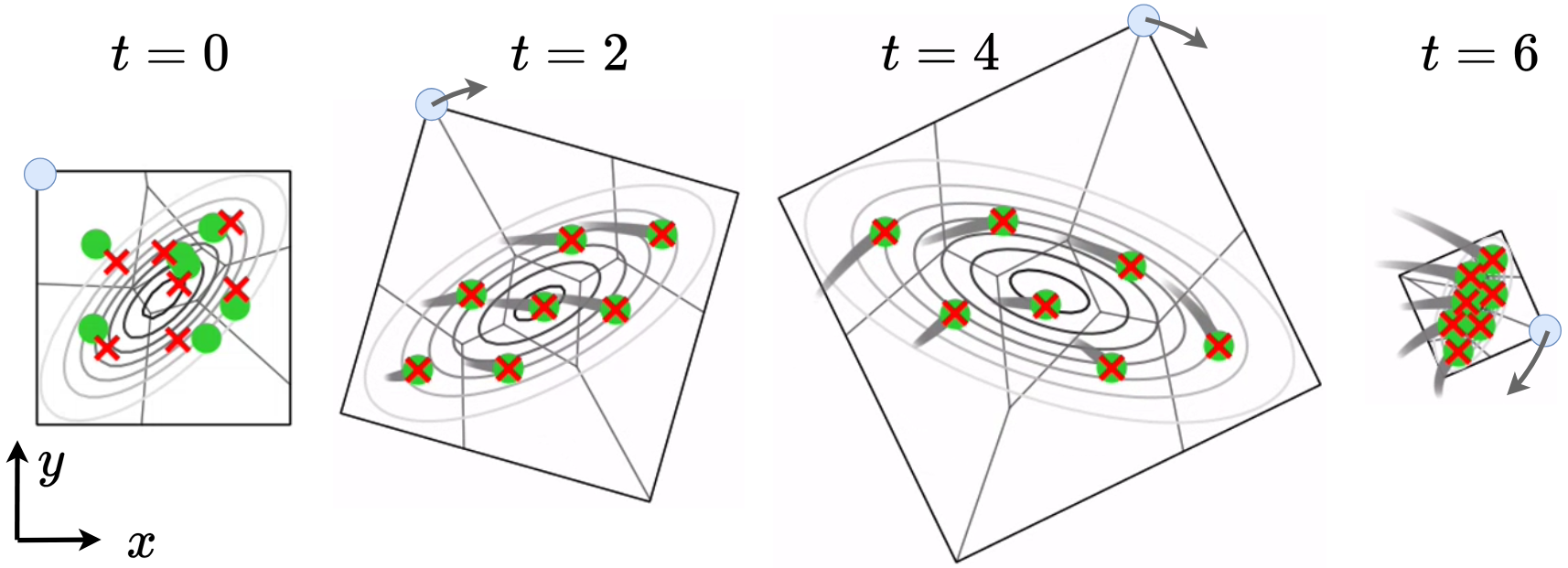}
    \caption{Agent formation under $u^i[z_t, \dot{z}_t]$ at $t \in \{0, 2, 4, 6\}$. Circles and streaks denote agent positions and recent agent trajectories. Contours denote $\phi_t$ and crosses and lines denote transformed partition centroids and boundaries. Note, $\Omega_t$ translates along the $x$-axis while rotating clockwise. }
    \label{fig:time_panel}
    \vspace{-5mm}
\end{figure}

\begin{figure}
    \centering
    \includegraphics[width=0.95\linewidth]{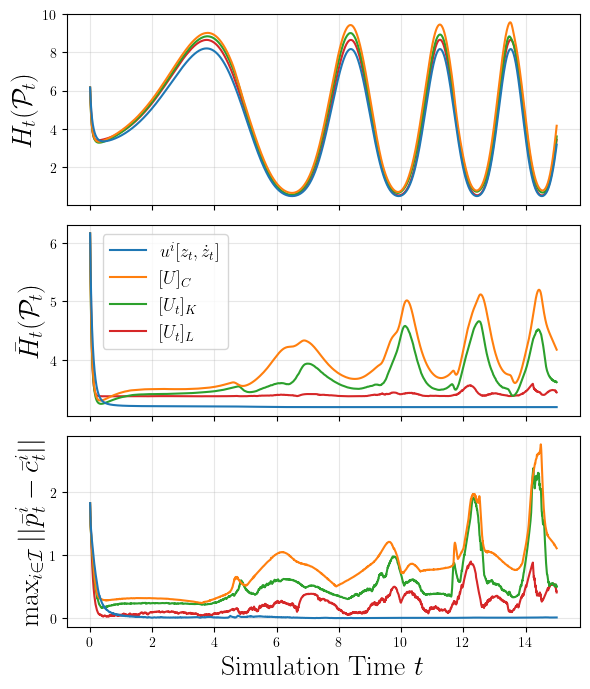}
    \caption{Coverage metrics along trajectories generated by coverage control laws: $u^i[z_t, \dot{z}_t]$ (diffeomorphic \eqref{eq:formation_controller}), $U_C$ (decentralised, static \cite{cortes2004coverage}), $[U_t]_K$ (decentralised, time-varying \cite{keene_acc}), $[U_t]_L$ (centralised, time-varying \cite{lee_multirobot_2015}).}
    \label{fig:loc_cost_plots}
\end{figure}

\begin{figure}
    \centering
    \includegraphics[width=1\linewidth]{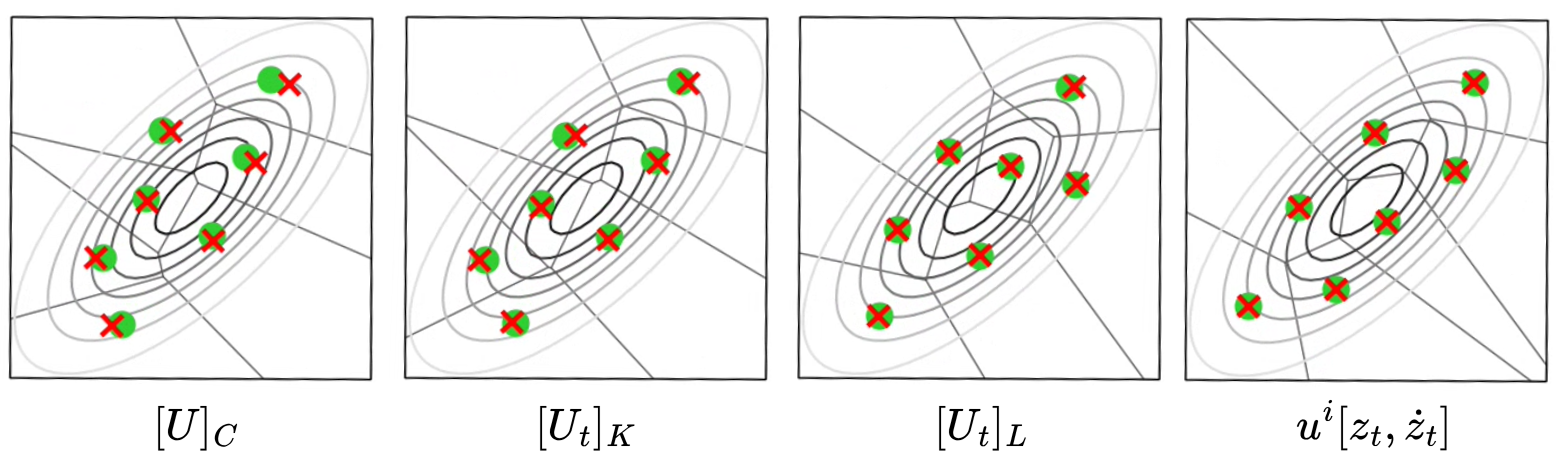}
    \caption{Formations at $t=13.4$.}
    \label{fig:controller_formations}
    \vspace{-3mm}
\end{figure}

Figure~\ref{fig:time_panel} shows simulated agent trajectories for $t \in [0, 6]$ under $u^i[z,\dot{z}]$, demonstrating effective formation 
over $t \mapsto (\Omega_t, \phi_t)$ despite significant relative motion. 
Figure~\ref{fig:loc_cost_plots} compares controllers across three metrics: the time-varying locational cost ($H_t$, \eqref{eq:H_t(P)}), diffeomorphic locational cost ($\nom{H}_t$, \eqref{eq:diffeo_loc_cost}), and maximum nominal-domain centroid-tracking error $\max_{i \in \mathcal{I}}\|\nom{p}^i - c^i[\nEnv,\nPhi](\nom{P})\|$. Figure~\ref{fig:controller_formations} presents agent formation snapshots for all controllers at $t = 13.4$.

\looseness=-1
The diffeomorphic coverage controller consistently outperforms all controllers across every metric. 
The decentralised controllers $U_C$ and $[U_t]_K$ demonstrate coverage degradation over the simulation as region-density velocity increases, reflected in large centroid-tracking errors and uneven formations in Figure~\ref{fig:controller_formations}. 
The centralised controller $[U_t]_L$, while less sensitive to region-density velocity than $U_C$ and $[U_t]_K$, converges to a suboptimal formation relative to $u^i[z,\dot{z}]$, resulting in consistently inferior performance. 
For the simulated class of region-density trajectory, the diffeomorphic coverage controller therefore achieves performance comparable to the centralised controller under a decentralised architecture.

\subsection{General Diffeomorphic Coverage Control}
\begin{figure}
    \centering
    \includegraphics[width=0.95\linewidth]{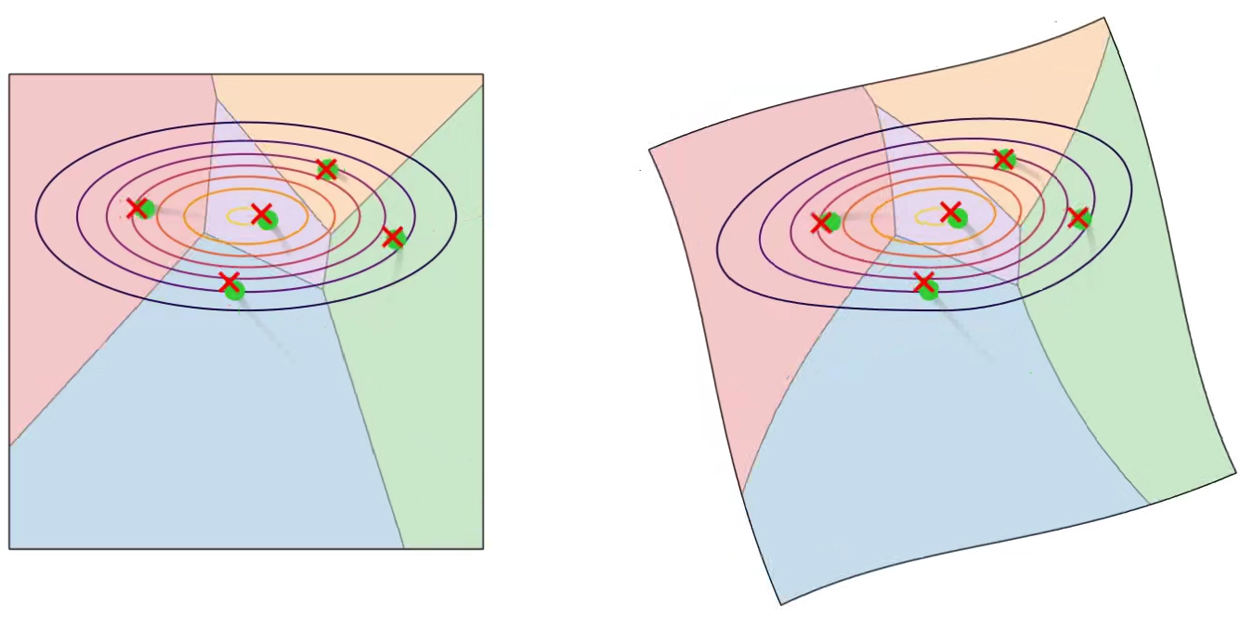}
    \includegraphics[width=0.95\linewidth]{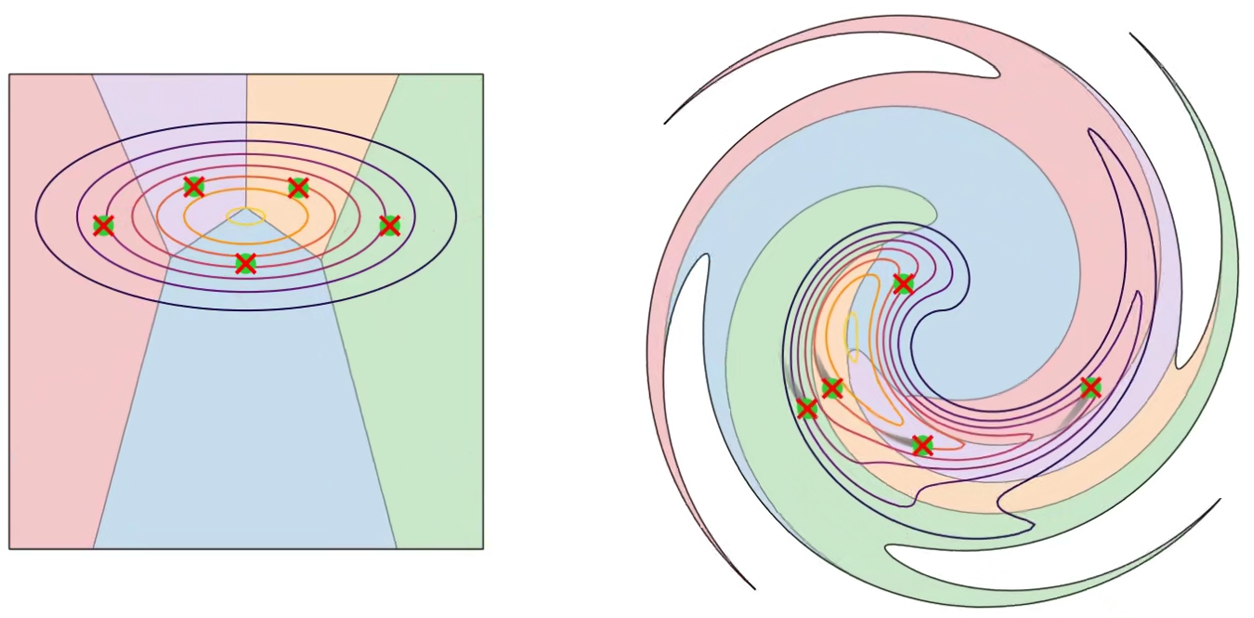}
    \caption{Coverage in nominal (left) and physical (right) domains at $t=1$ (top) and $t = 9$ (bottom).}
    \label{fig:diffeomorphic_region}
    \vspace{-3mm}
\end{figure}

\looseness=-1
Diffeomorphic coverage control over a non-affine diffeomorphism is also simulated. In this example, a multi-variate Gaussian and square coverage region evolve according to the non-uniform radially-dependent rotation
\begin{align*}
    \Psi_t(q) \doteq R(\theta_t(q))q, \quad \theta_t(q) \doteq 0.05t\norm{q}_2.
\end{align*}
Figure \ref{fig:diffeomorphic_region} displays a stable agent dispersion under \eqref{eq:diffeomorphic_controller} in the nominal and physical domains for $t\in\{1, 9\}$. Moreover, curved partition boundaries in Figure \ref{fig:diffeomorphic_region} emphasise the metric structure induced by $\Psi$ as discussed in Section \ref{sec:induced_metric_structure}. Significant boundary curvature is due to large distortion induced by $\Psi$, in contrast to the distortion-minimising cartograms constructed in \cite{Leonard_cartogram}. Furthermore, similar to \cite{Xu_nonconvex}, Figure \ref{fig:diffeomorphic_region} illustrates that $\Omega_t$ need not be convex under diffeomorphic coverage control which only requires convexity in $\nEnv$. 

\section{Conclusion}
\label{sec:Conclusion}
\looseness=-1
This paper addressed diffeomorphic coverage control, a structured class of time-varying coverage control in which region-density evolution is parameterised by a time-dependent diffeomorphism acting on a static region-density pair. 
Continuous- and discrete-time controllers were proposed that solve a related static coverage problem in a time-invariant nominal domain and transform the resultant control action into the physical domain.
The proposed laws are decentralised and asymptotically stable with respect to the diffeomorphic locational cost, a time-varying locational cost defined under a non-uniform metric. 
Simulations demonstrate controller effectiveness in a formation control task under an affine diffeomorphism, and under a non-uniform diffeomorphism.
Despite its decentralised architecture, the proposed controller achieves performance comparable to an equivalent centralised law.

\bibliography{biblio}  
\bibliographystyle{ieeetr}

\appendix
\subsection{Proofs of Lemma \ref{lem:continuous_equiv_systems} and Theorem \ref{theorem:continuous_time_stability}}
\label{append:cont_proofs}
\begin{proof}[Proof of Lemma 
\ref{lem:continuous_equiv_systems}]
    \looseness=-1
    Fix $ \Psi \in \bdiff{\nom{\mathcal{A}} }{\R^d}$, $x_0 \in \Psi_0(\nom{\mathcal{A}})$. Existence of $t \mapsto \xi_t[x_0]$ satisfying \eqref{eq:induced_dynamics} holds by inspection and the Picard-Lindel\"of theorem \cite{khalil2002nonlinear}, noting $\nom{U}$ is time-invariant and $\Psi \in \mathscr{C}^1(\mathcal{T}\times\mathbb{R}^d;\mathbb{R}^d)$. Moreover, boundedness of $\nom{U}$ over $\nom{\mathcal{A}}$ gives boundedness of $U_t[\Psi, \nom{U}](\Psi_t(\nom{x}))$ over $(t, \nom{x}) \in \CalT \times \nom{\mathcal{A}}$. 
    Define $\nom{\zeta}_{(\cdot)} \doteq \Psi_{(\cdot)}^{-1}(\xi_{(\cdot)}[x_0])$. Abusing notation with $\xi_{(\cdot)} = \xi_{(\cdot)}[x_0]$ and applying the chain rule gives
    \begin{align}
        \label{eq:P_tilde_deriv}
        \textstyle \tderiv{}{t} \nom{\zeta}_t =  \partial_t \Psi_t^{-1} \left(\xi_t \right) + D_x \Psi_t^{-1} \left(\xi_t \right) \textstyle\tderiv{}{t}\xi_t.
    \end{align}
    By $U_t[\Psi, \nom{U}]$ in \eqref{eq:induced_dynamics} and the definition of $\xi_t$ above, expressing $\textstyle\tderiv{}{t}\xi_t$ in terms of $\nom{\xi}_t$ gives
    \begin{align*}
        \textstyle\tderiv{}{t}\xi_t & = U_t[\Psi, \nom{U}](\xi_t) \\
        & = \partial_t \Psi_t \left(\Psi_t^{-1}(\xi_t)\right) +  D_x \Psi_t[\Psi_t^{-1}(\xi_t)] \nom{U} \left( \Psi_t^{-1}(\xi_t)\right)\\
         & = \partial_t \Psi_t (\nom{\zeta}_t) +  D_x \Psi_t[\Psi_t^{-1}(\xi_t)] \nom{U} ( \nom{\zeta}_t).
    \end{align*}
    Substituting $\textstyle\tderiv{}{t}\xi_t$ and the identities \eqref{eq:diff_id1}, \eqref{eq:diff_id2} into \eqref{eq:P_tilde_deriv} gives
    \begin{align*}
        \textstyle\tderiv{}{t}\nom{\zeta}_t & = \partial_t \Psi_t^{-1} \left(\xi_t \right) + D_x \Psi_t^{-1} \left(\xi_t \right) \partial_t \Psi_t (\nom{\zeta}_t) \\
        & \qquad + D_x \Psi_t^{-1} \left(\xi_t \right) \left[
        D_x \Psi_t[\Psi_t^{-1}(\xi_t)\right] \nom{U}( \nom{\zeta}_t)  \\
        & = \nom{U} ( \nom{\zeta}_t).
    \end{align*}
    \looseness=-1
    By definition, $\nom{\zeta}_0 = \Psi_0^{-1}(x_0) \in \nom{\mathcal{A}}$. Therefore, $t \mapsto \nom{\zeta}_{t}$ satisfies \eqref{eq:nominal_sys} and is a nominal system trajectory with $\nom{x}_0 \doteq \Psi_0^{-1}(x_0)$. Hence, $\Psi_{(\cdot)}^{-1}(\xi_{(\cdot)}[x_0])=\nom{\zeta}_{(\cdot)} = \nom{\xi}_{(\cdot)}[\nom{x}_0]$ is unique and $\Psi_t^{-1}(\xi_t[x_0]) \in \nom{A}$ for all $t \in \CalT$ by Definition \ref{def:NominalSystem}. The equivalence in \eqref{eq:equivalence} follows by invertibility of $\Psi_t$.
\end{proof}

\begin{proof}[Proof of Theorem 
\ref{theorem:continuous_time_stability}]
    Fix $x_0 \in \Psi_0(\nom{A}), \ \nom{x}_0 \doteq \Psi^{-1}_0(x_0) \in \nom{\mathcal{A}}$. By Lemma \ref{lem:continuous_equiv_systems}, $t \mapsto \xi_t[x_0]$ satisfying \eqref{eq:induced_dynamics} exists, is unique, and satisfies \eqref{eq:equivalence} such that $\Psi_t^{-1}(\xi_t[x_0]) = \nom{\xi}_t[\nom{x}_0]$ for all $t \in \CalT$. By an abuse of notation let $\nom{\xi}_{(\cdot)} \doteq \nom{\xi}_{(\cdot)}[\nom{x}_0]$, then
    \begin{align*}
        \textstyle\tderiv{}{t}V(\nom{\xi}_t) & = \nabla_{x}V(\nom{\xi}_t)^\top \textstyle \tderiv{}{t} \nom{\xi}_t = \nabla_{x}V(\nom{\xi}_t)^\top \nom{U}(\nom{\xi}_t) \leq 0,
    \end{align*}
    for all $t \in \CalT$. Hence, as $V(\nom{x}) \geq 0$ for all $\nom{x} \in \nom{\mathcal{A}}$, $t \mapsto \nom{\xi}_t$ converges to the largest invariant subset in $M[V, \nom{U}]$ by LaSalle's Invariance Principle \cite{khalil2002nonlinear}, i.e. $\lim_{t\to\infty}\textrm{dist}(\nom \xi_t[x_0], \eta[\nom{U}])=0$. 
    The theorem assertion then follows by the induced metric equivalence in Lemma \ref{cor:Metric_Equiv} noting $\Psi \in \bdiff{\nom{\mathcal{A}} }{\R^d}$. 
\end{proof}

\subsection{Diffeomorphic Lloyd Descent}
\label{sec:append_diffeom_lloyd}
For $\nom{\mathcal{A}} \subset \R^d$, define $\kbdiff{\nom{\mathcal{A}}}{\R^d}$ as the set of maps $\hat{\Psi}:\N \times \R^d \to \R^d$ such that $\hat{\Psi}_k=\Psi_{\tau(k)}$, where $\Psi\in \bdiff{\nom{\mathcal{A}}}{\R^d}$ and $\tau:\N\to\mathcal T$ is strictly monotone. 
We study discrete time-varying coverage control when $(\hat{\Omega}_k, \hat{\phi}_k) = \Gamma[\hat{\Psi}_k](\nEnv, \nPhi)$ as per \eqref{eq:diffeo_region_density} for all $k \in \N$ under $\hat{\Psi}\in\kbdiff{\nom{\mathcal{A}}}{\R^d}$. Discrete-time analogues of the nominal and physical systems from Section \ref{sec:General_Stab_section} are defined as follows. 

\begin{definition}[Discrete-time Regular Nominal System]
    \label{def:discrete_regular_nominal}
    The map $\nom{T}: \R^d \to \R^d$ generates a discrete nominal system over $\nom{\mathcal{A}} \subset \R^d$ if $\nom{T}(x)$ is well defined and $\nom{T}(\nom{x}) \in \nom{\mathcal{A}}$ for all $x \in \nom{\mathcal{A}}$.
\end{definition}

\begin{lemma}[Discrete-time Equivalent Physical System]
\label{lem:discrete_time_equiv}
Let $\nom{T}$ generate a discrete-time regular nominal system over $\nom{\mathcal{A}} \subset \R^d$ per Definition \ref{def:discrete_regular_nominal}. Given $\hat{\Psi} \in \kbdiff{\nom{\mathcal{A}} }{\R^d}$, $x_0 \in \hat{\Psi}_0(\nom{\mathcal{A}})$, let $\{\xi_k[x_0] \}_{k \in \N}$ denote the sequence satisfying
\begin{equation}
    \label{eq:discrete_controller}
    \begin{aligned}
        \xi_{k+1}[x_0] = T_{k}[\hat{\Psi}, \nom{T}]\left( \xi_{k}\right) , \qquad \xi_0[x_0] = x_0,\\
    T_{k}[\hat{\Psi}, \nom{T}]\left( x\right) \doteq \hat{\Psi}_{k+1} \circ \nom{T} \circ \hat{\Psi}_k^{-1} \left(x \right),
    \end{aligned}
\end{equation}
for all $k \in \N$. Likewise, let $\{\nom{\xi}_k[\nom{x}_0] \}_{k \in \N}, \  \nom{x}_0 \doteq \hat{\Psi}_0^{-1}(x_0)$ denote the sequence satisfying $\nom{\xi}_{k+1}[\nom{x}_0] = \nom{T}\left( \nom{x}_{k}\right), $ $\nom{\xi}_0[\nom{x}_0] = \nom{x}_0$. Then, $\xi_k \in \hat{\Psi}_k(\nom{\mathcal{A}})$ for all $k \in \N$ and
\begin{align*}
    \xi_{k}[x_0] = \hat{\Psi}_k(\nom{\xi}_k[\hat{\Psi}_0^{-1}(x_0)]), \qquad \forall k \in \N.
\end{align*}
\end{lemma}
\begin{proof}
    Fix $\hat{\Psi} \in \kbdiff{\nom{\mathcal{A}} }{\R^d}$, $x_0 \in \hat{\Psi}_0(\nom{\mathcal{A}})$.
    Define $\{\nom{\zeta}_k\}_{k \in \N}$ such that $\nom{\zeta}_k \doteq \Psi^{-1}_k(\xi_k[x_0])$ for all $k \in \N$. Then, by \eqref{eq:discrete_controller}
    \begin{align*}
        \nom{\zeta}_{k+1} = \Psi^{-1}_{k+1}(\xi_{k+1}[x_0]) = \nom{T} \left(\hat{\Psi}_k^{-1} \left( \xi_{k}[x_0]\right)\right) = \nom{T}(\nom{\zeta}_{k}).
    \end{align*}
    Therefore, as $\nom{\zeta}_0 = \nom{x}_0$, $\hat{\Psi}_{k}^{-1}(\xi_k[x_0]) =\nom{\zeta}_k = \nom{\xi}_k[\nom{x}_0]$ for all $k \in \N$ and the assertion follows by invertibility of $\hat{\Psi}^{-1}_k$.
\end{proof}

The discrete-time dynamics in \eqref{eq:discrete_controller} illustrate the simplicity of the discrete-time equivalent system. The composed structure maps inputs from the physical to nominal domain ($\hat{\Psi}^{-1}_{k}$), applies the nominal dynamics ($\nom{T}$) in the nominal domain, and maps the result back to the updated physical domain ($\hat{\Psi}_{k+1}$). Stability of the discrete-time equivalent follows by Theorem \ref{theorem:discrete_time_stability}. 

\begin{theorem}
\label{theorem:discrete_time_stability}
Let $\nom{T}$ generate a discrete-time regular nominal system over $\nom{\mathcal{A}} \subset \R^d$ per Definition \ref{def:discrete_regular_nominal} and define $\Gamma^*(\nom{T}) \doteq \{\nom{x} \in \nom{\mathcal{A}}  : \nom{T}(\nom{x}) = \nom{x} \}$. 
Given $\hat{\Psi} \in \kbdiff{\nom{\mathcal{A}} }{\R^d}$, $x_0 \in \hat{\Psi}_0(\nom{\mathcal{A}})$, denote the sequence in \eqref{eq:discrete_controller} as $\{\xi_k[x_0] \}_{k \in \N}$.
Suppose there exists a continuous map $\nom Z:\R^d \to \Rpos$ such that $\nom Z(\nom{T}(\nom x))\leq \nom Z( \nom x)$ for all $\nom x \in \nom{\mathcal{A}}$ with strict inequality iff $\nom x \in \Gamma^*(\nom{T})$.
Then $\textrm{dist}(\xi_k, \hat{\Psi}_k(\Gamma^*(\nom{T}))) \to 0$ as $k \to \infty$. Furthermore, if $\Gamma^*(\nom{T})$ is finite then there exists $\nom{x}^* \in \Gamma^*(\nom{T})$ such that $\textrm{dist}(\xi_k, \hat{\Psi}_k(\nom x^*)) \to 0$ as $k \to \infty$.
\end{theorem}
\begin{proof}
    \looseness=-1
    Fix $\hat{\Psi} \in \kbdiff{\nom{\mathcal{A}} }{\R^d}$, $x_0 \in \hat{\Psi}_0(\nom{\mathcal{A}})$. 
    By Lemma \ref{lem:discrete_time_equiv}, there exists a trajectory $\{\nom{\xi}_k\}_{k \in \N}$ such that $\xi_k = \hat{\Psi}_k(\nom{\xi}_k)$ with $\nom{\xi}_{k+1} = \nom{T}(\nom{\xi}_k)$ for all $k \in \N$. 
    Therefore $\nom{\xi}_0 = \Psi^{-1}_0(x_0) \in \nom{\mathcal{A}}$ and $\nom{\xi}_k \in \nom{\mathcal A}$ for all $k \in \N$ by Definition \ref{def:discrete_regular_nominal}. 
    Let $\nom{x}^*$ denote an accumulation point of $\{\nom{\xi}_k\}_{k \in \N}$. By the Global Convergence Theorem \cite[Lemma 1.3]{cortes2004coverage} and descent function $\nom Z$, $\nom{x}^* \in \Gamma^*(\nom{T})$ and $\textrm{dist}(\nom{\xi}_k, \Gamma^*(\nom{T})) \to 0$ as $k \to \infty$. 
    The theorem assertion follows by the induced metric equivalence in Lemma \ref{cor:Metric_Equiv} noting $\hat{\Psi}\in\kbdiff{\nom{\mathcal{A}} }{\R^d}$. The final assertion follows from \cite[Proposition 1.4]{cortes2004coverage}.
\end{proof}

Discrete-time coverage control is achieved via the well-known Lloyd descent \cite{lloyd_least_1982}. For a time-invariant region-density pair $(\nEnv, \nPhi)$ the Lloyd algorithm is defined by the map $\nom{T}[\nEnv, \nPhi]: \nEnv^n \backslash \mathcal{S} \to \nEnv^n \backslash \mathcal{S}$ given by
\begin{equation}
    \label{eq:lloyd_map}
    \begin{aligned}
    \nom{T}[\nEnv, \nPhi](P) & \doteq \left[c^i[\nEnv, \nPhi](P)\right]_{i \in \mathcal{I}}.
    \end{aligned}
\end{equation}

Time-varying discrete-time coverage in the physical domain then achieved via \textit{diffeomorphic Lloyd descent} via specialising \eqref{eq:discrete_controller} with the discrete nominal system generated by \eqref{eq:lloyd_map}. Asymptotic stability follows by the following corollary of Theorem \ref{theorem:discrete_time_stability}.
\begin{corollary}[Diffeomorphic Lloyd Descent]
    \label{prop:discrete_coverage}
    Suppose $k \mapsto (\Omega_{k}, \phi_{k})$ is diffeomorphic with respect to $(\nEnv, \nPhi)$ and $\hat{\Psi} \in \kbdiff{\nEnv}{\R^2}$ as per Definition \ref{def:diffeomorphic_region-density}.
    Let $\{P_k\}_{k \in \N}$ denote the sequence starting at $P_0 \in (\hat{\Psi}_0(\nEnv))^n \backslash \mathcal{S}$ such that $P_{k+1} = T_k[\hat{\Psi}, \nom{T}[\nEnv, \nPhi]](P_k)$ as per \eqref{eq:discrete_controller} and \eqref{eq:lloyd_map} for all $k \in \N$. 
    Then, $P_k \in \hat{\Psi}_k(\nom{\mathcal{A}})$ for all $ k \in \N$ and $\textrm{dist}(P_k, \hat{\Psi}_k({\mathcal{C}[\nEnv, \nPhi]})) \to 0$ as $k \to \infty$. 
    Furthermore, if $\mathcal{C}[\nEnv, \nPhi]$ is finite, then there exists $C^* \in \mathcal{C}[\nEnv, \nPhi]$ such that $\textrm{dist}(P_k, \hat{\Psi}_k(C^*)) \to 0$ as $k \to \infty$.
\end{corollary}
\begin{proof}
    \looseness=-1
    Fix $P_0 \in (\hat{\Psi}_0(\nEnv))^n \backslash \mathcal{S}$. The map $\nom{T}[\nEnv, \nPhi]$ generates a discrete-time regular nominal system over $\nEnv^n \backslash \mathcal{S}$ \cite{cortes2004coverage}. 
    The result then follows by Theorem \ref{theorem:discrete_time_stability} using the nominal-domain locational cost, $\mathcal{H}[\nEnv, \nPhi]$ as per \eqref{eq:calH(P)}, as a candidate function. By \cite{cortes2004coverage}, $\mathcal{H}[\nEnv, \nPhi]$, is decrescent under the algorithm $\nom{T}[\nEnv, \nPhi]$. In addition, $\Gamma^*(\nom{T}[\nEnv, \nPhi]) = \mathcal{C}[\nEnv, \nPhi]$. Therefore, $\textrm{dist}(P_k, \hat{\Psi}_k({\mathcal{C}[\nEnv, \nPhi]})) \to 0$ as $k \to \infty$. The final assertion follows by Theorem \ref{theorem:discrete_time_stability}.
\end{proof}
\end{document}